\documentclass[lettersize,journal]{IEEEtran}
\usepackage{amsmath,amsfonts}
\usepackage{algorithmic}
\usepackage{cite}
\usepackage{array}
\usepackage[caption=false,font=normalsize,labelfont=sf,textfont=sf]{subfig}
\usepackage{textcomp}
\usepackage{stfloats}
\usepackage{url}
\usepackage{verbatim}
\usepackage{graphicx}
\usepackage{xcolor}
\usepackage{multirow}
\def\BibTeX{{\rm B\kern-.05em{\sc i\kern-.025em b}\kern-.08em
		T\kern-.1667em\lower.7ex\hbox{E}\kern-.125emX}}
\usepackage{balance}
\begin{document}
	\title{{Intelligent Rotatable Antenna for Integrated Sensing, Communication, and Computation: Challenges and Opportunities}}
\author{ 
 Xue Xiong,	Beixiong Zheng,~\IEEEmembership{Senior Member,~IEEE}, Wen Wu,~\IEEEmembership{Senior Member,~IEEE}, Weihua Zhu,\\ 
 Miaowen Wen,~\IEEEmembership{Senior Member,~IEEE}, Shaoe Lin, and Yong Zeng,~\IEEEmembership{Fellow, IEEE} 
}
	
	
	\maketitle 
	
\begin{abstract}
Integrated sensing, communication, and computation (ISCC) has emerged as a promising paradigm for enabling intelligent services in future sixth-generation (6G) networks. 
However, existing ISCC systems based on fixed-antenna architectures inherently lack spatial adaptability to cope with the signal degradation and dynamic environmental conditions. 
Recently, non-fixed flexible antenna architectures, such as fluid antenna system (FAS), movable antenna (MA), and pinching antenna, have garnered significant interest. Among them, intelligent rotatable antenna (IRA) is an emerging technology that offers significant potential to better support the comprehensive services of target sensing, data transmission, and edge computing. 
This article investigates a novel IRA-enabled ISCC framework to enhance received signal strength, wider coverage, and spatial adaptability to dynamic wireless environments by flexibly adjusting the boresight of directional antennas.
Building upon this, we introduce the fundamentals of IRA technology and explore IRA's benefits for improving system performance while providing potential task-oriented applications. 
Then, we discuss the main design issues and provide solutions for implementing IRA-based ISCC systems.
Finally, experimental results are provided to demonstrate the great potential of IRA-enabled ISCC system, thus paving the way for more robust and efficient future wireless networks.
\end{abstract}
	
%
	
\section{Introduction}
The advent of the sixth-generation (6G) wireless network signifies a fundamental transformation in mobile communications, which are expected to support intelligent and mission-critical services such as autonomous driving, virtual/augmented reality, and smart manufacturing. 
These emerging applications call for the fusion of physical, biological, and cyber worlds, which inherently involves three core functionalities, including environmental sensing to acquire information, wireless communication for information sharing, and computation to process data and make intelligent decisions \cite{Shen2022Holistic}.
However, conventional wireless architectures typically treat sensing, communication, and computation as separate modules, leading to inefficient resource utilization, increased latency, and inadequate intelligence.
This has thus motivated research efforts to explore the integration and coordination gains achieved by combining these individual functionalities, giving rise to frameworks such as integrated sensing and communication (ISAC), integrated communication and computation, and integrated sensing and computation.
While some progress has been made in partial combinations, they fall short of meeting the stringent performance requirements for real-time intelligent tasks, such as ultra-high decision accuracy and ultra-low latency.
Consequently, integrated sensing, communication, and computation (ISCC) has emerged as a unified framework, aiming to comprehensively integrate all three functionalities within a single system to improve resource coordination and reduce end-to-end task delay \cite{zhu2023pushing}. 
Existing ISCC systems with fixed-antenna architectures generally improve performance by expanding the scale of antenna arrays, which, however, leads to increased hardware and manufacturing complexity, higher radio-frequency (RF) energy consumption, and greater signal processing overhead. 
Additionally, the lack of dynamic adaptability to channel variations often results in performance degradation, especially in environments characterized by dense users/targets distribution and high mobility.

The development of flexible antenna architectures, such as movable antenna (MA) and fluid antenna system (FAS), has marked a significant advancement in wireless technologies \cite{zhu2023modeling,wong2020fluid,zheng2024flexible,Zhu2024MAMagazine}.
Specifically, by dynamically reconfiguring the positions and/or shapes of the antenna elements, MA and FAS can exploit the spatial diversity within a predefined space to create favorable channel conditions. 
This introduces additional spatial degrees of freedom (DoFs) to enhance wireless communication and/or sensing performance. 
Unlike traditional multi-antenna systems that rely on numerous fixed antennas and RF chains to boost capacity, MA and FAS can achieve comparable gains with much fewer antennas.
As a result, they significantly reduce the hardware complexity and energy consumption.
To further unlock additional spatial DoFs, six-dimensional MA (6DMA) incorporates antenna reconfigurability by enabling both three-dimensional (3D) position and rotation adjustments \cite{Shao20256DMA}.
While this additional flexibility further enhances the adaptability to wireless environment, it also increases design complexity and implementation cost due to the intricate geometric and mechanical constraints for antenna movement.

Recently, intelligent rotatable antenna (IRA) has been proposed as a novel and effective solution to enhance spatial DoFs while maintaining a compact aperture and low hardware cost. 
Generally, IRA can flexibly rotate the radiation pattern towards desired directions by independently adjusting the 3D boresight direction of each antenna via mechanical or electronic means \cite{zheng2025rotatable,zheng2025magazine}.
Owing to its directional control flexibility and spatial reconfigurability, IRA offers a new opportunity to enhance ISCC networks by reshaping wireless channel conditions, thereby boosting the transmission/data offloading rate and expanding wireless coverage. 
One of the promising advantages of IRA lies in its capability to conduct real-time directional beam adaptation, enabling the system to steer beams toward sensing targets, communication users, or edge computing nodes in accordance with network requirements. 
This not only enhances the transmission link quality and beam alignment accuracy, but also facilitates the dynamic allocation of antenna resources based on the spatial distribution of targets, users, and edge servers.
Building upon this, IRA technology offers a powerful mechanism for spatial task decoupling, allowing different antenna resources to be independently oriented to support simultaneous environmental sensing, data transmission, and computation offloading in distinct spatial directions. 
Such decoupling is essential for ISCC networks where tasks often compete for limited spectral and spatial resources.
Furthermore, IRA significantly contributes to task-level performance optimization across all three ISCC domains. For sensing, the unique beam tracking and scanning capabilities of IRA enhance both sensing resolution and coverage. 
For communication, IRA can enhance signal quality and suppress interference by 
leveraging flexible beamforming and exploiting spatial multiplexing gains. 
For computation, by providing multi-view data collection, enhancing offloading link reliability, and enabling adaptive task-aware scheduling, IRA can significantly enhance the computation efficiency and responsiveness.

\begin{figure*}[!t]
	\centering
	\includegraphics[width=1\textwidth]{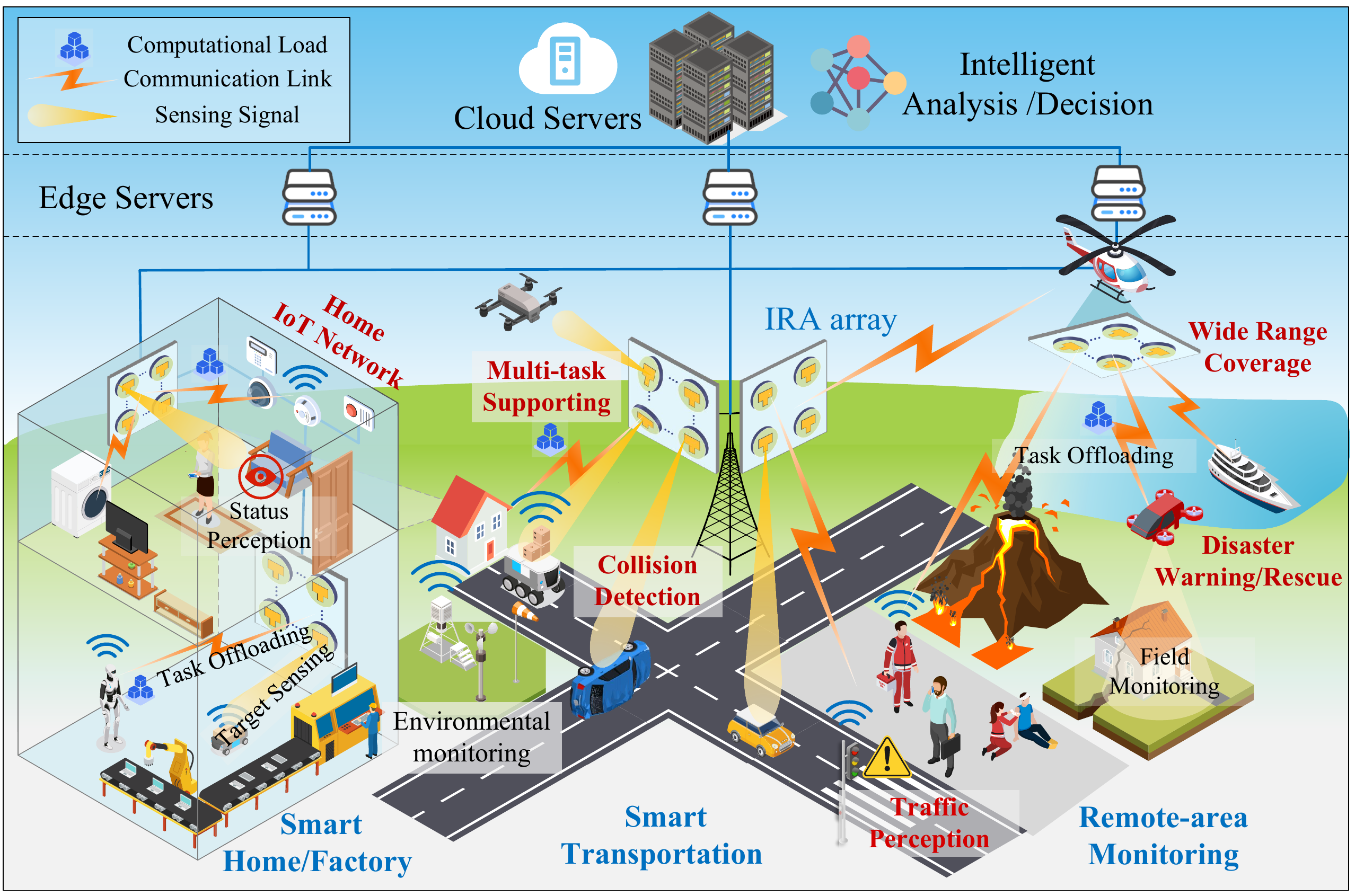}
	\caption{Typical scenarios of IRA-enabled ISCC systems.}
	\label{Scenario_ISCC}
\end{figure*}
In Fig. \ref{Scenario_ISCC}, we envision various applications in future IRA-enabled ISCC networks, such as smart transportation, smart homes and factories, and remote-area monitoring.
In this overarching framework, IRA serves as a promising enabler for future ISCC systems in various scenarios owing to the following key advantages.
\begin{itemize}
	\item \textbf{Directional beam adaptation:} IRA enables fine-grained 3D beam steering by adjusting antenna orientations/boresights, allowing the system to dynamically align beams toward sensing targets, communication users, or edge servers based on the task-specific demands.
	\item \textbf{Enhanced spatial multiplexing:} The ability to reorient individual antennas allows for improved spatial separation among users or targets, supporting simultaneous environmental sensing, data transmission, and computation offloading with minimal mutual interference.
	\item \textbf{Interference mitigation and suppression:} IRA can steer nulls toward strong interferers or isolate functional domains spatially, providing an effective mechanism for mitigating intra- and inter-task interference in ISCC systems.
	\item \textbf{Improved sensing resolution and coverage:} By directing beams with higher angular precision, IRA increases spatial resolution and extends coverage in desired directions, thus enhancing the reliability and accuracy of sensing tasks.
	\item \textbf{Task-aware resource coordination:} The spatial flexibility of IRA enables task-driven allocation of spatial resources, thus facilitating efficient scheduling of directional beams for optimized ISCC systems.
\end{itemize}

The aforementioned benefits of IRA motivate us to further explore and unlock the full potential of IRA in ISCC networks. Therefore, this article provides a comprehensive review of IRA-enabled ISCC systems, focusing on their fundamental principles and performance advantages, diverse task-oriented applications, and practical design challenges. 
Additionally, through experimental results, we illustrate how IRA technology can substantially enhance ISCC performance.	

\section{Fundamentals of IRA-enabled ISCC}

\subsection{Fundamentals of IRA}
\begin{figure*}[!t]
	\centering
	\includegraphics[width=1\textwidth]{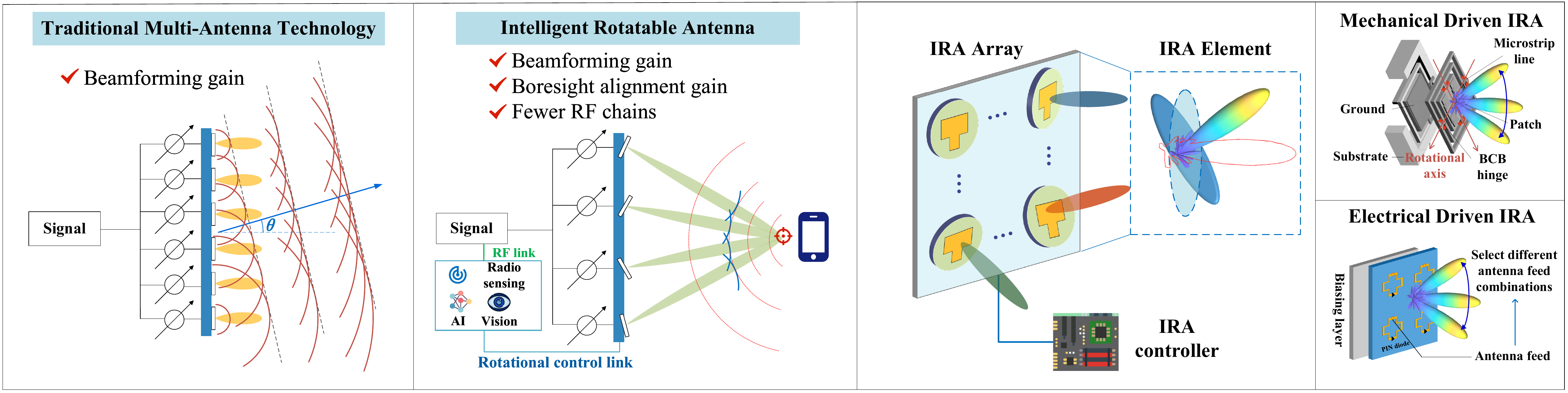}
	\caption{Architectures of IRA.}
	\label{RA}
\end{figure*}
A typical IRA architecture is illustrated in Fig. \ref{RA}. In addition to the communication module, which includes communication components such as RF chains and baseband processing units in conventional fixed antenna architectures, IRA incorporates a dedicated control module for rotating the boresight direction of each antenna or antenna array. 
As such, IRA enables both beamforming gain and boresight alignment gain to concentrate signal energy toward intended targets,  thus achieving spatial beam focusing with significantly fewer RF chains compared to traditional multi-antenna technology. 
Depending on the implementation requirements, this boresight adjustment can be achieved through mechanical mechanisms, electronic mechanisms, or a hybrid combination of both.

The mechanically driven IRA typically employs servo-motor-controlled platforms \cite{dai2025rotatable,dai2025demo} or micro-electromechanical systems (MEMS) \cite{baek2003v} to physically rotate the antenna orientation, thereby adjusting the boresight direction in 3D space. This approach provides a wide range of boresight adjustment and high precision in beam alignment, but may suffer from relatively low response speeds.
Alternatively, the electronically driven IRA achieves boresight adjustment by configuring multi-feed antenna or adjusting parasitic elements, thereby enabling rapid changes in the main lobe direction without altering the physical orientation of the antenna.
This is accomplished by activating different feed ports or electronically tuning loads such as varactor diodes or PIN diodes on parasitic elements \cite{morshed2023beam,boyarsky2021electronically}. 
Electrical control method can quickly and finely adjust boresight direction, but has a relatively limited adjustment range.
Both mechanisms typically provide discrete and predefined radiation directions.
To enable continuous and precise control, more advanced tunable materials like liquid crystals can be utilized.
To combine the strengths of both, hybrid IRA architecture that integrates mechanical
and electronic control can also be employed, thereby balancing wide-angle adaptability with rapid radiation pattern reconfiguration.

In addition to hardware architecture, a key aspect of IRA is the characterization of how its radiation pattern is manipulated during operation.
The radiation pattern of IRA is determined by its antenna design, which defines critical parameters such as mainlobe gain, beamwidth, and sidelobe levels. 
These intrinsic parameters remain constant, while the boresight direction can be dynamically adjusted through mechanical or electronic rotation.
Unlike MA that shifts the pattern via physical translation, IRA rotates the entire radiation pattern, thus enabling precise directional alignment toward desired direction.
The flexibility of rotation depends on the hardware design and environmental constraints. High-resolution adjustments require precise actuators or multiple excitation states, which may increase hardware complexity. 
Moreover, the rotation range is limited by control module and physical surroundings.
By carefully optimizing control accuracy and rotation range, IRA can enhance signal directivity, suppress undesired interference, and improve overall system performance.
This capability makes IRA particularly well-suited for ISCC systems that require adaptive and concurrent multi-task execution across dynamic environments.

\subsection{IRA-Enabled ISCC}
The primary characteristic of the ISCC framework lies in its ability to deeply integrate wireless sensing with wireless communication technology, while leveraging widely distributed computing capabilities for auxiliary processing \cite{wen2024survey}. 
This integration paradigm is of paramount significance for supporting intelligent servers in future wireless networks.
However, existing ISCC systems that rely on fixed antenna architectures often necessitate a substantial number of antennas and/or RF chains, leading to prohibitively high signal processing overhead.
Additionally, the performance of ISCC is passively constrained by unfavorable propagation conditions due to dynamic environments and unpredictable blockages. 
To address these challenges, IRA provides a promising solution by enabling real-time directional control and spatial task separation, thereby significantly enhancing environmental adaptability, spatial resource efficiency, and system robustness under dynamic conditions. 
The core idea of IRA-enabled ISCC system is to flexibly reconfigure IRA's boresight direction for creating more favorable channel conditions and increasing the probability of line-of-sight transmission (including communication and task offloading) as well as enhancing sensing capacity.
In addition, IRA can be directed towards different sensing regions to enhance sensing resolution and/or aligned with different communication users and edge servers to improve link quality.

\begin{figure}[!t]
	\centering
	\includegraphics[width=3.4in]{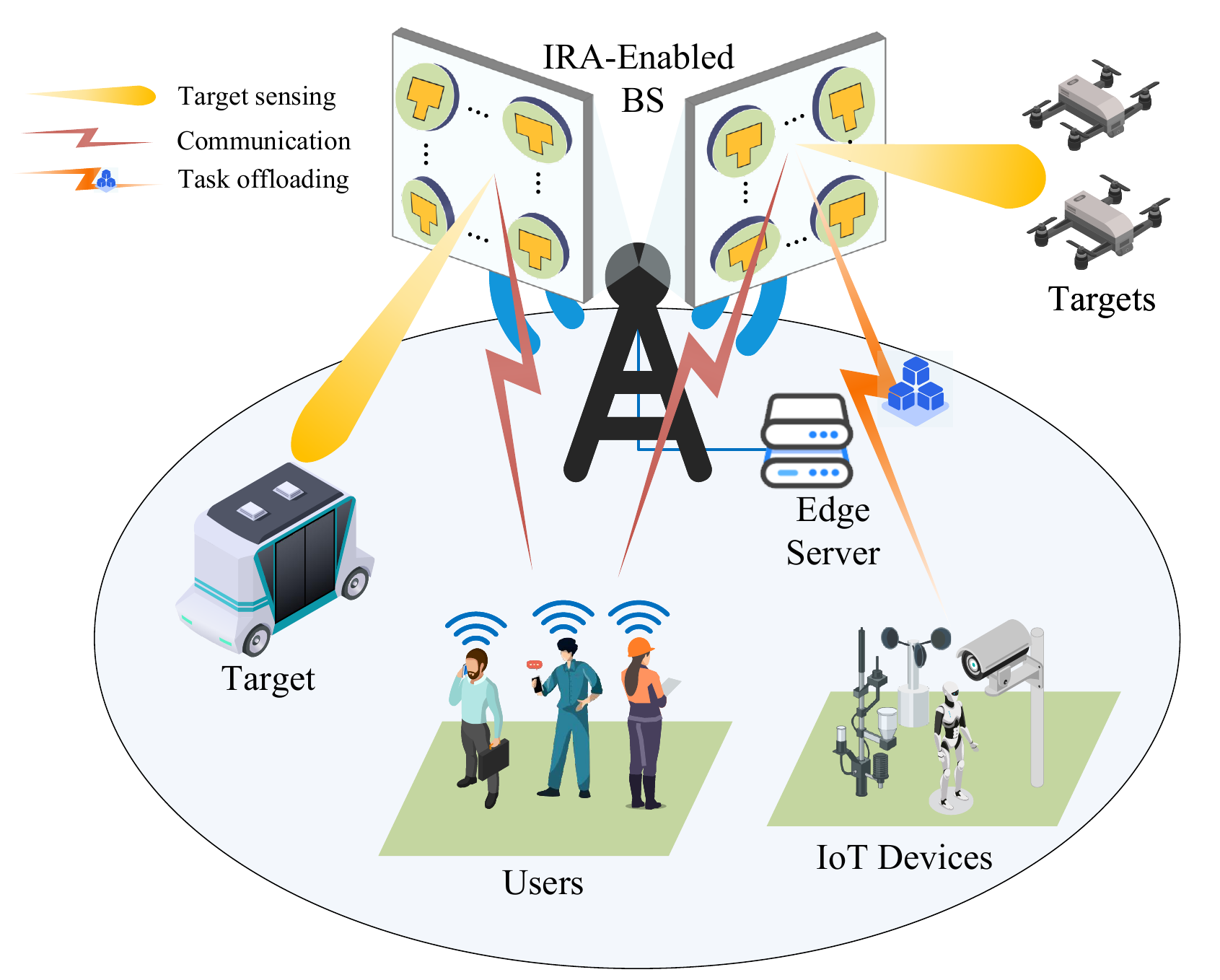}
	\caption{IRA-enabled ISCC system.}
	\label{System}
\end{figure}

A typical IRA-enabled ISCC system is shown in {Fig. \ref{System}}, where the triple-functional BS equipped with an edge server and multiple IRAs aims to conduct sensing tasks and provide communication and computing services.  
Each Internet-of-Things (IoT) device has its own computation task. Due to the limited computation resource at the IoT devices, the computation tasks need to be offloaded to the edge server located at the BS via wireless links.
Meanwhile, the BS needs to sense numerous targets and communicate with multiple users. 
For this purpose, the IRA deployed at the BS transmits ISAC signals towards sensing targets and communication users, collects the echo signals, and then feeds them back to the edge server for further processing.    
By enabling independent 3D boresight control across different antennas in the array, IRA can be directed toward distinct sensing targets to enhance angular resolution and/or aligned with specific communication users to improve the quality of transmission links.
Simultaneously, by strategically leveraging the orientation adjustment of IRAs, the task offloading channel conditions and computational offloading efficiency can be significantly improved.

Generally, IRA-enhanced ISCC systems can achieve superior performance by utilizing much fewer antennas to exploit additional spatial DoFs.
This framework significantly reduces the signal processing overhead and system complexity compared to traditional ISCC systems with fixed antenna architectures, but also enhances the overall efficiency of resource utilization.
By appropriately allocating IRA elements based on the requirements of communication, sensing, and computation, as well as the spatial channel distribution of users, targets, and IoT devices, the ISCC framework can effectively support multi-functional tasks.  
In typical configurations, an IRA-enabled ISCC system performs three main functions: sensing, communication, and computation. 
For sensing tasks, the overall sensing framework consists of two phases, i.e., the echo signal collection phase at the BS and the signal processing phase at the edge server. 
The former is responsible for sensing the surroundings and collecting the echo signals, while the latter processes these signals to determine the status of targets.
Through diverse angular observations enabled by antenna boresight adjustment, the IRA-enabled BS can extract multiple target features such as direction, velocity, size, and orientation. 
For communication tasks, the IRA-enabled BS transmits dual-functional signals toward specific users and targets. 
The IRA can adjust the antenna's main lobe to suppress or nullify interference while maintaining high gains toward intended users. 
For computation tasks, the offloading process for each device can be divided into two main phases: 1) the transmission phase, where each IoT device transmits its data to the BS via allocated time slots; and 2) the computation phase, during which the edge server processes the task. 
By appropriately adjusting the configurable orientations of IRA, the system can enhance the offloading channel conditions and thereby improve computational efficiency. 

\subsection {Task-oriented ISCC systems empowered by IRA}
In diverse intelligent applications, the sensing, communication, and computation tasks possess distinct requirements and priorities. 
This subsection classifies the IRA-enabled ISCC systems according to their dominant task objective and elaborates on how IRA can be optimally leveraged in each context to enhance system performance.
\begin{itemize}
	\item \textbf{Sensing-oriented ISCC:} 
	In applications such as autonomous driving and smart surveillance, real-time and high-resolution sensing is of paramount importance.
	Sensing-oriented ISCC systems prioritize the allocation of wireless and computational resources to maximize sensing accuracy and/or coverage, with communication and computation serving as supportive functions.
	IRA offers substantial advantages in this context through the ability to dynamically steer beams and perform spatial scanning. 
	Unlike fixed-antenna architecture, IRA array can continuously adjust the antenna boresight directions to sequentially or concurrently probe multiple spatial regions, thus enabling the detection of closely spaced targets with much fewer antennas \cite{zheng2025magazine}. This flexibility enhances angular resolution, mitigates signal overlap, and improves spatial separability. Furthermore, by capturing signals from diverse angular perspectives, IRA facilitates the extraction of rich target features such as direction, velocity, size, and orientation, which are crucial for robust multidimensional perception in dynamic environments.
	\item \textbf{Communication-oriented ISCC:} 
	In scenarios such as industrial IoT and emergency communications, reliable and high-capacity data transmission becomes the primary system objective. 
	Communication-oriented ISCC systems allocate more spectrum and computational resources for robust connectivity and high throughput, while sensing and computation support the tasks including channel state inference and beam tracking.
	In this context, IRA enhances communication performance by enabling real-time and fine-grained directional control. 
	Unlike fixed or digitally steered arrays, IRA array can physically rotate antenna's boresight to align with user locations or beamforming directions, thereby enhancing signal strength and reducing interference \cite{zheng2025rotatable}. 
	This element-level beam steering is particularly beneficial for users in dynamic or obstructed environments, maintaining link reliability with lower latency and power consumption. Moreover, the ability to reconfigure radiation patterns adaptively supports user-centric coverage, spatial multiplexing, and interference mitigation in large-scale user access scenarios.
	\item \textbf{Computation-oriented ISCC:} 
	In applications such as collaborative edge intelligence and smart manufacturing, real-time processing of massive data streams is essential. Computation-oriented ISCC systems focus on optimizing resource allocation to enable efficient data offloading and distributed computation, while sensing and communication act as enablers for accurate data acquisition and reliable data exchange.
	In this regards, IRA plays a crucial role in enhancing computation performance by improving the quality and directionality of transmission links from distributed devices to edge servers. By steering beams toward computing nodes, IRA establishes stronger channels, thereby reducing transmission delay and energy consumption during task offloading. 
	Furthermore, IRA can dynamically prioritize high-priority nodes based on task urgency or mobility, facilitating efficient scheduling of computational resources. 
	For distributed learning or inference tasks, IRA assists in aggregating multi-source data with minimal collision or interference, ensuring more synchronized and reliable model updates. 
\end{itemize}

\section{Main design issues and solutions}
While IRA technology shows great potential in ISCC systems for enhancing multi-function coordination and resources utilization efficiency, new design issues also need to be addressed for IRA-enabled ISCC systems. 
In the following, we elaborate on the main issues in model and estimation, multiple access and interference cancellation, IRA deployment and boresight direction optimization, as well as performance trade-off.
In addition, we provide forward-looking solutions for future exploration.
\subsection{Modeling and Estimation}
Accurate modeling and parameter estimation are essential for unlocking the full potential of IRA-enabled ISCC systems. As IRA enables real-time adjustment of antenna boresight directions, it also introduces dynamic spatial reconfigurability and orientation-dependent variations in wireless channels. This will change traditional modeling approaches and impose new modeling requirements. 
Specifically, in IRA-enabled systems, the wireless channel impulse response depends not only on transceiver locations but also on antenna orientations. 
A key challenge is quantifying how dynamic antenna orientations affect wireless channels characteristics. 
To address this, a pair of rotation angles can be introduced to describe the 3D boresight direction for each IRA, allowing the channel model to be incorporated with these angles \cite{zheng2025rotatable}.
Furthermore, in practical ISCC applications, sensing tasks often operate in the near-field region, while communication and computation tasks are typically confined to the far-field region. 
Thus, a unified framework is required to jointly account for near-field spherical wavefronts and far-field planar propagation, thereby enabling modeling of heterogeneous propagation characteristics across different functional domains. 
Moreover, the ISCC systems are inherently heterogeneous, involving multi-user communication, multi-target sensing, and task-driven computation offloading.
The coexistence of these functionalities introduces spatial coupling
effects, which complicate spatial resource modeling and
challenge traditional isolation-based design approaches.
As such, new task-aware model should be developed to characterize the spatial correlations and interdependencies among sensing, communication, and computing tasks.

In terms of parameter estimation, extracting accurate and real-time channel state information (CSI) under dynamic antenna orientations poses new challenges. 
Unlike FAS and MA/6DMA \cite{zheng2024flexible}, IRA maintains fixed antenna positions during operation, preserving the transceiver geometry and simplifying spatial modeling. 
This structural stability allows for estimating fewer environmental parameters, similar to fixed-antenna systems. 
However, orientation-dependent channel dynamics necessitate fine-grained estimation of direction-sensitive parameters. 
Moreover, since IRA can collect data from multiple spatial perspectives, effectively fusing multi-view sensory data for environmental inference becomes crucial, particularly in complex or cluttered environments. To tackle these challenges, knowledge-assisted estimation methods, such as Kalman filtering or mobility-aware prediction, can be utilized to track time- and orientation-varying channels. Low-rank and sparse recovery techniques help reduce estimation overhead by leveraging the channel structured sparsity. Additionally, emerging deep learning (DL)-based models trained on multi-view data can enhance estimation robustness under non-ideal conditions, although the absence of  standard datasets for IRA system remains an unresolved issue.

\subsection{Multiple Access and Interference Management}
The integration of sensing, communication, and computation tasks over a shared wireless infrastructure inevitably introduces intricate interference and access management challenges \cite{wen2024survey}.
While the IRA-based architecture allows multiple tasks and users to dynamically share spatial and spectral resources via rotatable beam control, this flexibility also increases the risk of cross-domain interference.
First, due to the spatial convergence of multiple tasks on a common IRA array, different beams for sensing, communication, and computation may partially overlap in direction, particularly under limited spatial DoFs in predefined region. 
This directional coupling may lead to mutual-task interference, especially in dynamic scenarios with time-varying priorities or user mobility. 
Second, when multiple users or targets are served simultaneously, multi-user interference (MUI) becomes pronounced.
Although IRA provides enhanced directionality, obtaining precise beam separation becomes increasingly challenging in high-density environments with limited spatial resolution.
Third, computation-oriented systems often involve bursty uplink transmissions due to massive task offloading. 
If these task flows are not adequately aligned in both spatial and temporal domains, they may lead to channel congestion and inter-stream interference, ultimately degrading the overall ISCC performance. 
Furthermore, the dynamic nature of IRA boresight adjustment adds more control complexity, rendering traditional interference mitigation techniques such as fixed beamforming/precoding ineffective due to the rapid changes in channel conditions with antenna orientation.

To address these challenges, several promising directions can be further explored. 
Task-aware IRA scheduling allows for dynamic allocation of antenna resources by accounting for task priorities, spatial requirements, and multi-functional beam separability. 
Joint directional beam scheduling techniques can be developed to design orthogonal beam pattern across the sensing and communication functions, thereby reducing cross-domain leakage. 
Furthermore, environment-aware interference prediction frameworks based on directional interference graphs can be utilized to proactively predict and mitigate spatial overlaps before they occur. 
Lastly, adaptive learning-based algorithms, such as federated learning or reinforcement learning, can be employed to optimize multi-user access strategies and manage dynamic interference under time-varying environmental and traffic conditions.

\subsection{IRA Configuration and Boresight Optimization}
Determining the optimal configuration and boresights of IRA array is critical for fully leveraging the advantages of IRA in ISCC systems. 
From an implementation perspective, the IRA-enabled system can consist of either a single element or multiple elements, depending on the task requirements and hardware limitations. 
A single-IRA setup provides structural simplicity and cost-effective control, enabling basic sensing-communication-computation operations through time-division multiplexing. 
However, its limited spatial DoFs restrict performance in dynamic or multi-task environments. 
In contrast, multi-IRA configurations enable spatially decoupled task execution, advanced beamforming, and effective interference mitigation. These benefits come with increased hardware complexity, higher control overhead, and greater energy consumption, particularly in systems employing antenna-level mechanical rotation.
Therefore, it is essential to strike a balance among the number of antenna elements, rotation scale (array-level or antenna-level), and control mechanisms (mechanical or electronic-based means).
On the other hand, the environment-aware deployment strategies are indispensable. 
Compact arrays with electronically driven IRA elements are more suitable for size-constrained platforms such as IoT devices, while distributed IRA elements  provide coverage and macro-diversity benefits in wide-area or blockage-prone scenarios. 
By customizing the configuration to specific tasks, we can achieve both spatial efficiency and robust performance.

Once the IRA configuration is determined, finding the optimal boresight direction of each antenna in an adaptive manner is crucial for maximizing the performance gain of IRA-enabled ISCC systems. 
However, this problem is challenging to address owing to the inherent complexity and interdependence among multiple objectives.
Depending on system setups and task priorities, two main optimization paradigms can be employed. 
The first paradigm involves jointly optimizing the boresight directions alongside other resource blocks (e.g., time slots, bandwidth, and power) to maximize the overall system utility. This multi-objective optimization aims to balance sensing accuracy, communication throughput, and computational efficiency within a unified framework. The second paradigm focuses on task-oriented optimization, which prioritizes specific task requirements, such as aligning beams for sensing, reducing latency for computation, or maintaining communication links.
The corresponding optimization problems are generally non-convex and NP-hard, particularly when the number of IRAs is extremely large. 
Another challenge arises from the practical limitation of discrete boresight selections for each IRA. 
To tackle this issue, one practical approach involves relaxing the discrete constraints to solve the problem using continuous boresight direction values and subsequently quantizing the solutions to their nearest discrete levels. 
Alternatively, the low-complexity heuristic algorithms, such as greedy or iterative search methods, can efficiently explore the discrete solution space.
Another critical aspect is that the boresight optimization heavily relies on available CSI, which is challenging to obtain perfectly in practice. 
The robust optimization strategies can be employed to design antenna orientations/boresight directions under imperfect CSI conditions, thereby enhancing the system robustness against estimation errors.

\subsection{Performance Trade-off Among Sensing, Communication, and Computing}
In IRA-enabled ISCC systems, sensing, communication, and computation are no longer isolated modules but are deeply integrated into a unified wireless infrastructure.
While this integration enhances system efficiency, it also introduces inherent performance trade-offs due to the shared utilization of spectral, spatial, and computational resources. Specifically, the sensing necessitates wide angular coverage and high-precision real-time environmental awareness; communication demands stable link quality and high throughput; and computation benefits from low-latency task offloading and reliable data delivery. 
However, these objectives often conflict when simultaneously conducted within resource-constrained systems.
One of the key challenges lies in the coupling of spatially directional resources. 
For instance, the boresight direction of IRA that prefers a high effective gain for a communication link may not align with the optimal direction for target detection in sensing tasks.
Similarly, allocating bandwidth to enhance sensing resolution might reduce the available capacity for computational task offloading. 
This results in a multi-task competition for resources such as time, angle, power, and frequency.
Moreover, practical ISCC applications with time-varying user positions, dynamic task demands, and channel states, which significantly complicates the joint optimization of directional beam control and resource scheduling. 

To address these challenges, several strategies can be adopted. First, task-prioritized coordination frameworks can be developed to dynamically allocate resources in alignment with mission-critical objectives. 
For instance, communication quality feedback can guide sensing beam selection, while sensing data can facilitate channel state prediction and beam tracking. Second, optimization framework incorporating multi-objective criteria, such as Pareto-efficient frontiers, can be utilized to systematically characterize and navigate the trade-off space. Lastly, the learning-based methods, such as reinforcement learning or graph neural networks, can model long-term trade-off dynamics and learn adaptive policies under high uncertainty. Overall, the capability of IRA to reconfigure spatial resources in real time provides new opportunities for balancing performance across sensing, communication, and computing. 
In the future, ISCC systems will benefit from intelligent trade-off-aware designs that synergistically leverage spatial agility, environmental feedback, and multi-task coordination to optimize system-level performance.


%
%

\section{Experiment Results}
\begin{figure*}[!t]
	\centering
    \includegraphics[width=1\textwidth]{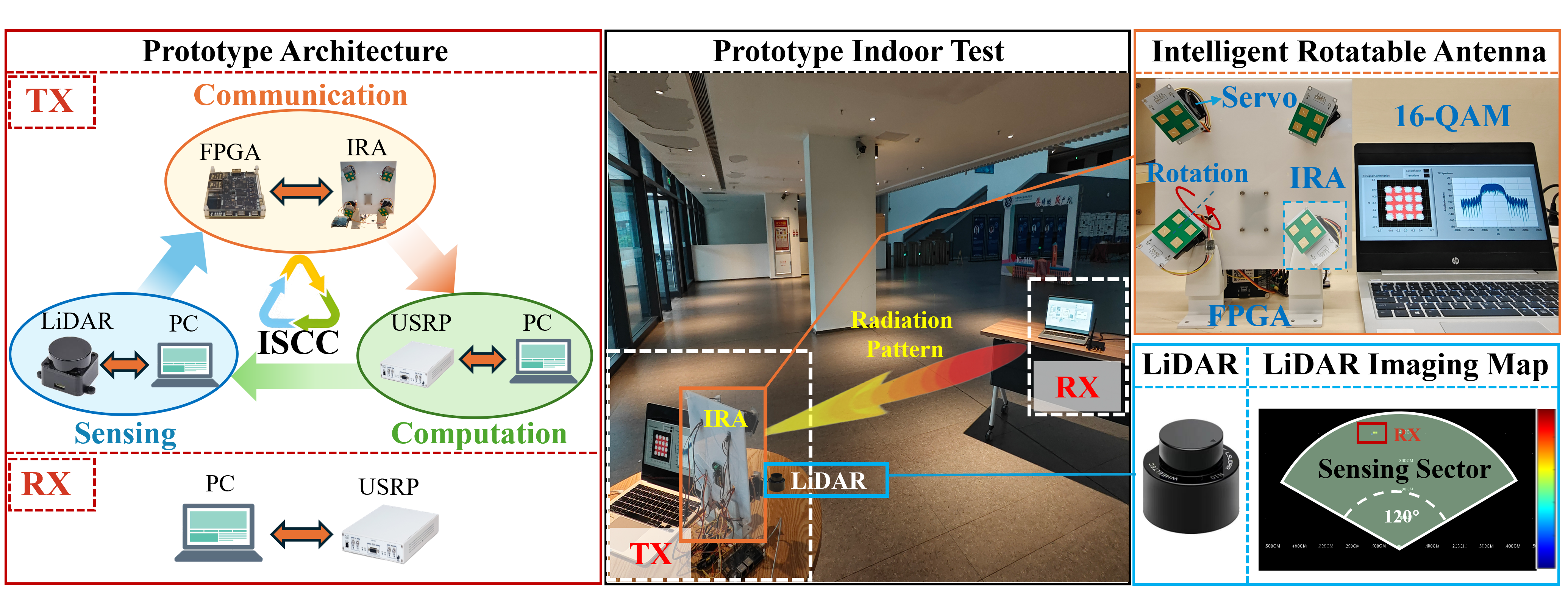}
	\caption{Prototype of IRA-enabled ISCC system.}
	\label{test_scenario1}
\end{figure*}
In this section, we develop an initial prototype and present the experimental result to validate the performance advantages of IRA-enabled ISCC system.
As shown in  Fig. \ref{test_scenario1}, the prototype consists of two main components: an IRA-enabled transmitter (TX) and a fixed-antenna receiver (RX), which operates in a typical indoor environment. 
Specifically, the TX incorporates a servo-controlled IRA, a LiDAR sensor for sensing environment, and a universal software radio peripherals (USRP)-based communication module.
At the RX side, a conventional USRP receiver is used to measure the received signal strength and decode the transmitted data. 
Besides, a personal computer (PC) is responsible for processing sensing data, analyzing received signals, and controlling beam steering. 

The IRA prototype system operates through a closed-loop ISCC process. 1) \textbf{Sensing:} The LiDAR mounted on the TX detects the position of the RX by scanning the surrounding environment. 
This sensing process estimates the angle-of-arrival (AoA) information, which is then used to assist subsequent beam alignment decisions. 
2) \textbf{Communication:} Based on the obtained AoA information, the IRA adjusts its boresight direction via a micro-controller (e.g., FPGA) and servo motor, physically rotating to align its radiation pattern toward the RX direction. 
The physical beam steering ensures the optimal signal alignment, thereby maximizing the received signal-to-noise (SNR) at the RX. 
3) \textbf{Computation:} All collected sensing data are transmitted to the PC.
This computing module processes multi-source inputs, evaluates system performance, and dynamically adjusts the IRA's orientation/boresight direction. 
It effectively closes the loop between sensing and actuation, enabling adaptive beam control according to the real-time environmental feedback.

\begin{figure}[!t]
	\centering
	\includegraphics[width=3.6in]{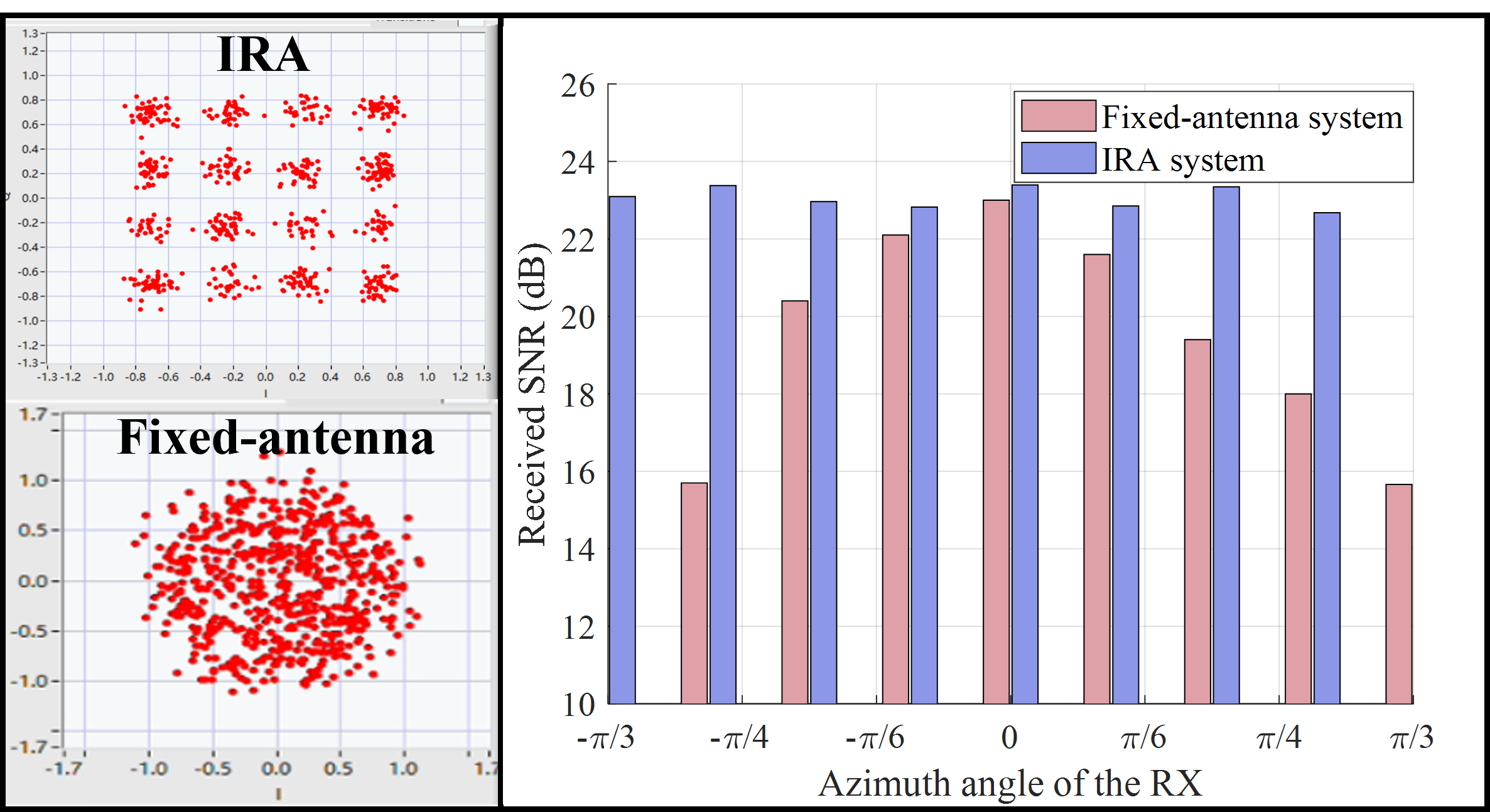}
	\caption{Constellation graphs and the received SNR versus azimuth angle of the RX.}
	\label{SNR_vs_Azimuth_angle}
\end{figure}

The settings for the IRA prototype system are specified as follows.
The IRA-equipped TX operates at a carrier frequency of 5.8 GHz with 16-quadrature amplitude modulation (QAM). The transmit power is set to 10 dBm, and the distance between the TX and RX is 4 m. Additionally, the bandwidth is configured to 100 kHz.
Fig. \ref{SNR_vs_Azimuth_angle} illustrates the constellation diagrams
of IRA and fixed-antenna systems, as well as the received SNR at the RX side versus its azimuth angle, with the zenith angle fixed at $\theta=0^{\circ}$. 
Compared to the fixed-antenna system, the constellation diagram of the IRA
system is clearer and more regularly distributed, indicating a
significant improvement in received signal quality.
Moreover, as the RX's azimuth angle varies from $-\frac{\pi}{3}$ to $\frac{\pi}{3}$, the IRA dynamically adjusts its boresight direction to track and align with the RX's direction, thus maintaining a stable received SNR. 
In contrast, for the fixed-antenna TX, the RX movement results in growing misalignment between the antenna boresight and the RX direction, thus leading to a significantly degradation in the received SNR. 
This result validates the effectiveness of IRA for enhancing the performance of ISCC systems.
The real-time sensing feedback guides directional beam adjustment, ensuring optimal link alignment and high signal quality.
Furthermore, the PC processor dynamically tunes IRA control based on real-time environmental and signal metrics, improving responsiveness to environmental changes.

\section{Conclusion}
In this article, we provided a comprehensive exploration of IRA-enabled ISCC systems, illustrating how IRA can unlock additional spatial DoFs to enhance the coordination across sensing, communication, and computation tasks.
We first presented the integration of IRA technology into the ISCC framework, emphasizing its fundamental principles and its advantages over traditional fixed-antenna techniques.
Then, we discussed the main design challenges, including modeling and estimation, multiple access and interference management, as well as IRA configuration and boresight optimization, while providing promising solutions as potential future research directions. 
Finally, experiments results have been presented to demonstrate the substantial performance gains achieved by IRA-enabled ISCC systems.
These findings highlight the potential of IRA as a cost-effective and scalable enabler for the next-generation intelligent wireless system.


\bibliographystyle{IEEEtran}
\bibliography{IRS_Stealth}

\section*{Biographies}
\noindent{{\bf Xue Xiong}
	(ftxuexiong@mail.scut.edu.cn) is with the School of Future Technology, South China University of Technology, Guangzhou, China, and also with the Frontier Research Center, Peng Cheng Laboratory, Shenzhen, China.
}
\\

\noindent{{\bf Beixiong Zheng}
	 (bxzheng@scut.edu.cn) is an Associate Professor with the School of Microelectronics, South China University of Technology, Guangzhou, China.
 }
\\

\noindent{{\bf Wen Wu}
	(wuw02@pcl.ac.cn) is an Associate Researcher with the Frontier Research Center, Peng Cheng Laboratory, Shenzhen, China. 
}
\\

\noindent{{\bf Weihua Zhu}
	(miweihuazhu@mail.scut.edu.cn) is with the School of Electronic and Information Engineering, South China University of Technology, Guangzhou, China.
}
\\

\noindent{{\bf Miaowen Wen}
	(eemwwen@scut.edu.cn) is a Professor with the School of Electronic and Information Engineering, South China University of Technology, Guangzhou, China.
}
\\

\noindent{{\bf Shaoe Lin}
	(linshaoe@oamail.gdufs.edu.cn) is with the School of Information Science and Technology, Guangdong University of Foreign Studies, Guangzhou, China.
}
\\

\noindent{{\bf Yong Zeng}
	(yong{\_}zeng@seu.edu.cn) is a Professor with the National Mobile Communications Research Laboratory, Southeast University, Nanjing, China, and also with the Purple Mountain Laboratories, Nanjing, China. 
}

\end{document}